\documentclass{aa}  
\usepackage{longtable}
\usepackage{graphicx}
\usepackage{subcaption}
\usepackage{natbib}
\bibpunct{(}{)}{;}{a}{}{,}

\usepackage{upgreek} 
\usepackage{xspace} 
\usepackage{color, colortbl} 

\newcommand{\ciii}{C\,{\sc iii}]\xspace}
\newcommand{\civ}{C\,{\sc iv}\xspace}
\newcommand{\nii}{[N\,{\sc ii}]\xspace}
\newcommand{\mgii}{\ion{Mg}{ii}\xspace}
\newcommand{\sii}{[S\,{\sc ii}]\xspace}
\newcommand{\neiii}{[Ne\,{\sc iii}]\xspace}
\newcommand{\oii}{[O\,{\sc ii}]\xspace}
\newcommand{\oiii}{[O\,{\sc iii}]\xspace}
\newcommand{\oi}{O\,{\sc i}\xspace}

\newcommand{\hei}{He\,{\sc i}\xspace}
\newcommand{\heii}{He\,{\sc ii}\xspace}
\newcommand{\hi}{H\,{\sc i}\xspace}
\newcommand{\lya}{Ly{$\rm \alpha$}\xspace}
\newcommand{\ha}{H{$\rm \alpha$}\xspace}
\newcommand{\hb}{H{$\rm \beta$}\xspace}
\newcommand{\hg}{H{$\rm \gamma$}\xspace}
\newcommand{\hd}{H{$\rm \delta$}\xspace}

\newcommand{\lrd}{J0647\_1045\xspace}

\begin{document}
\title{Deciphering the JWST spectrum of a `little red dot' at $z\sim4.53$: An obscured AGN and its star-forming host}


\titlerunning{Little red dot at $z\sim4.5$}

\author{Meghana Killi\inst{1,2,3}, Darach Watson\inst{1,2}, Gabriel Brammer\inst{1,2}, Conor McPartland\inst{1,2}, Jacqueline Antwi-Danso\inst{4}, Rosa Newshore\inst{5}, Dan Coe\inst{6,7,8}, Natalie Allen\inst{1,2}, Johan P. U. Fynbo\inst{1,2}, Katriona Gould\inst{1,2}, Kasper E. Heintz\inst{1,2}, Vadim Rusakov\inst{1,2}, Simone Vejlgaard\inst{1,2}}

\authorrunning{Killi et al.}

\institute{Cosmic Dawn Center (DAWN), Jagtvej 128, 2200 Copenhagen N, Denmark\\
        \email{meghana.killi@mail.udp.cl}     
        \and
        Niels Bohr Institute, University of Copenhagen, Lyngbyvej 2, 2100 Copenhagen Ø, Denmark 
        \and
        Instituto de Estudios Astrofísicos, Facultad de Ingeniería y Ciencias, Universidad Diego Portales, Av. Ejército 441, Santiago 8370191, Chile
        \and
        David A. Dunlap Department of Astronomy \& Astrophysics, University of Toronto, 50 St George Street, Toronto, ON M5S 3H4, Canada
        \and
        Department of Physics, Clark University Worcester, MA 01610-1477
        \and
        Space Telescope Science Institute (STScI), 3700 San Martin Drive, Baltimore, MD 21218, USA
        \and
        Association of Universities for Research in Astronomy (AURA) for the European Space Agency (ESA), STScI, Baltimore, MD, USA
        \and
        Center for Astrophysical Sciences, Department of Physics and Astronomy, The Johns Hopkins University, 3400 N Charles St. Baltimore, MD 21218, USA
        }

\date{Received ...; accepted ...}

\abstract
{\emph{JWST} has revealed a class of numerous, extremely compact sources, with rest-frame red optical/near-infrared (NIR) and blue ultraviolet (UV) colours, nicknamed ``little red dots''. We present one of the highest signal-to-noise ratio \emph{JWST} NIRSpec/PRISM spectra of a little red dot, \lrd at $z=4.5321\pm0.0001$, and examine its NIRCam morphology, to differentiate the origin of the UV and optical/NIR emission, and elucidate the nature of the little red dot phenomenon. \lrd is unresolved ($r_e \lesssim 0.17$\,kpc) in the three NIRCam long-wavelength filters, but significantly extended ($r_e = 0.45\pm0.06$\,kpc) in the three short-wavelength filters, indicating a red compact source in a blue star-forming galaxy. The spectral continuum shows a clear change in slope, from blue in the optical/UV, to red in the restframe optical/NIR, consistent with two distinct components, fit by power-laws with different attenuation: $A_V=0.54\pm0.01$ (UV) and $A_V=5.7\pm0.2$ (optical/NIR). Fitting the \ha line requires both broad (full-width at half maximum $\sim 4300\pm300$\,km\,s$^{-1}$) and narrow components, but none of the other emission lines, including \hb, show evidence of broadness. We calculate $A_V = 1.1\pm0.2$ from the Balmer decrement using narrow \ha and \hb, and $A_V > 4.1\pm0.2$ from broad \ha and upper limit on broad \hb, consistent with the blue and red continuum attenuation respectively. Based on single-epoch \ha linewidth, the mass of the central black hole is $8\pm1 \times 10^{8}$\,M$_\odot$. Our findings are consistent with a multi-component model, where the optical/NIR and broad lines arise from a highly obscured, spatially unresolved region, likely a relatively massive AGN, while the less obscured UV continuum and narrow lines arise, at least partly, from a small but spatially resolved star-forming host galaxy.}


\maketitle

\section{Introduction}
\label{sec:intro}

\emph{JWST} surveys of extragalactic fields have revealed abundant red, point or near-point sources \citep[e.g.][]{Barro2023ExtremelyAGNs, Maiolino2023JADES.Mighty, Harikane2023JWST/NIRSpecProperties, Greene2023UNCOVERz5}, sometimes referred to as ``little red dots'' \citep[LRDs;][]{Matthee2023EIGER.JWST}. These somewhat mysterious sources appear to be dominated by emission from active galactic nuclei (AGN) at $z\sim4-9$. Estimates of their supermassive black hole (SMBH) masses based on their modestly broad (1200--4000\,km\,s$^{-1}$) \ha and \hb emission lines, suggest a wide range of masses \citep[10$^5$--10$^9$\,M$_\odot$;][]{Furtak2023AShadows, Kocevski2023HiddenCEERS, Matthee2023EIGER.JWST, Maiolino2023JADES.Mighty, Greene2023UNCOVERz5, Larson2023AQuasars, Kokorev2023UNCOVER:8.50}. LRDs also exhibit an unusual ``V-shaped'' spectral energy distribution (SED), i.e.\ a spectrum with very red restframe optical colours and blue UV colours \citep[e.g.][]{Furtak2023AShadows, Barro2023ExtremelyAGNs, Fujimoto2023CEERSProperties, Greene2023UNCOVERz5}. The unusual SEDs suggest that LRDs might be something more exotic than simple type~1 AGN.

High-$z$ AGN provide a window into early SMBH formation and growth. SMBHs with masses estimated to be in excess of a billion solar masses have now been found as early as 600\,Myr after the Big Bang \citep[e.g.][]{Wang2021A7.642}. This poses a challenge to black hole growth scenarios involving not only the largest stellar mass black holes \citep{Ohkubo2009EvolutionCollapse, Chantavat2023TheStars} but also the hypothesised direct-collapse black holes \citep{Bromm2003FormationHoles, Trinca2022TheDawn, Natarajan2023FirstCollapse, Schneider2023Are9}. The discovery of LRDs, a population of lower mass SMBHs in the early universe that may be in a dust-obscured, rapid growth phase \citep[e.g.][]{Fujimoto2022ADawn}, can help alleviate the tension. With up to 100 times the number densities of UV-selected black holes \citep{Barro2023ExtremelyAGNs, Greene2023UNCOVERz5, Furtak2023AShadows}, LRDs are a significant population among early AGN.

While many LRDs in the literature have a compact morphology, several show a spatially extended component as well \citep{Harikane2023JWST/NIRSpecProperties}. Together with the unusual SED shape, this indicates that the contribution of the host galaxies to LRD spectra and morphology is non-negligible. LRDs therefore provide a unique opportunity to study not only the numbers and level of activity of early AGN, but also the host galaxies and their relation to the central AGN. This simultaneous view into early, young, obscured AGN and their hosts can tell us about the larger environment that SMBH growth took place in, and how this in turn affected the host, including properties such as star formation rate, spatial extent, dust obscuration, and gas composition and ionisation.

A downside is that as contribution from the host increases and dominates over that from the AGN, it becomes increasingly difficult to identify the source as an AGN \citep{Onoue2023AUniverse}. Several authors have pointed out that the broadness in Balmer emission lines is currently the only reliable method to detect these obscured, low-mass AGN \citep[e.g.][]{Kocevski2023HiddenCEERS, Matthee2023EIGER.JWST}. This detectability in turn has implications for their estimated number counts. It has thus become crucial to understand the nature of these objects, estimate the contribution of the AGN and host to their observed properties, and trace their evolution from the early Universe to the present.

In this paper, we study \lrd, a prototypical example of an LRD at $z=4.5321\pm0.0001$ (not including systematic uncertainty; see Sec.~\ref{sec:bestfit_spec}), with a low-resolution (but high signal-to-noise ratio; SNR) NIRSpec prism spectrum, and imaging in NIRCam bands from 1--5\,$\upmu$m. We consider the origin of the observed spectral features in an AGN or star-forming host galaxy, and use the spectral shape, line ratios, linewidths, and morphology to guide our conclusions.

We assume a cosmology based on the Planck 2018 data \citep{Aghanim2020PlanckParameters}, with $\rm H_0=67.66\,km\,s^{-1}\,Mpc^{-1}$ and $\Omega_M=0.31$. 
Stellar masses and star-formation rates are based on a Chabrier initial mass function \citep{Chabrier2003GalacticFunction}.

\section{Methods}
\subsection{Observations and Data}

\begin{figure}
    \centering
    \includegraphics[width=0.4\textwidth]{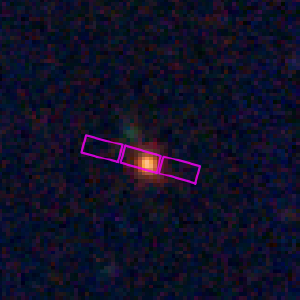}
    \caption[\lrd F115W image]{RGB image (created using the F115W, F277W, and F444W NIRCam filters) of \lrd, with the NIRSpec slits (magenta rectangles) overlaid. The image is centered on the coordinates of the source at RA=101.933406, Dec=70.198268, and is 2\,arcsec on a side.}
    \label{fig:rgb_slit}
\end{figure}


\begin{figure*}
    \centering
    \includegraphics[width=\textwidth]{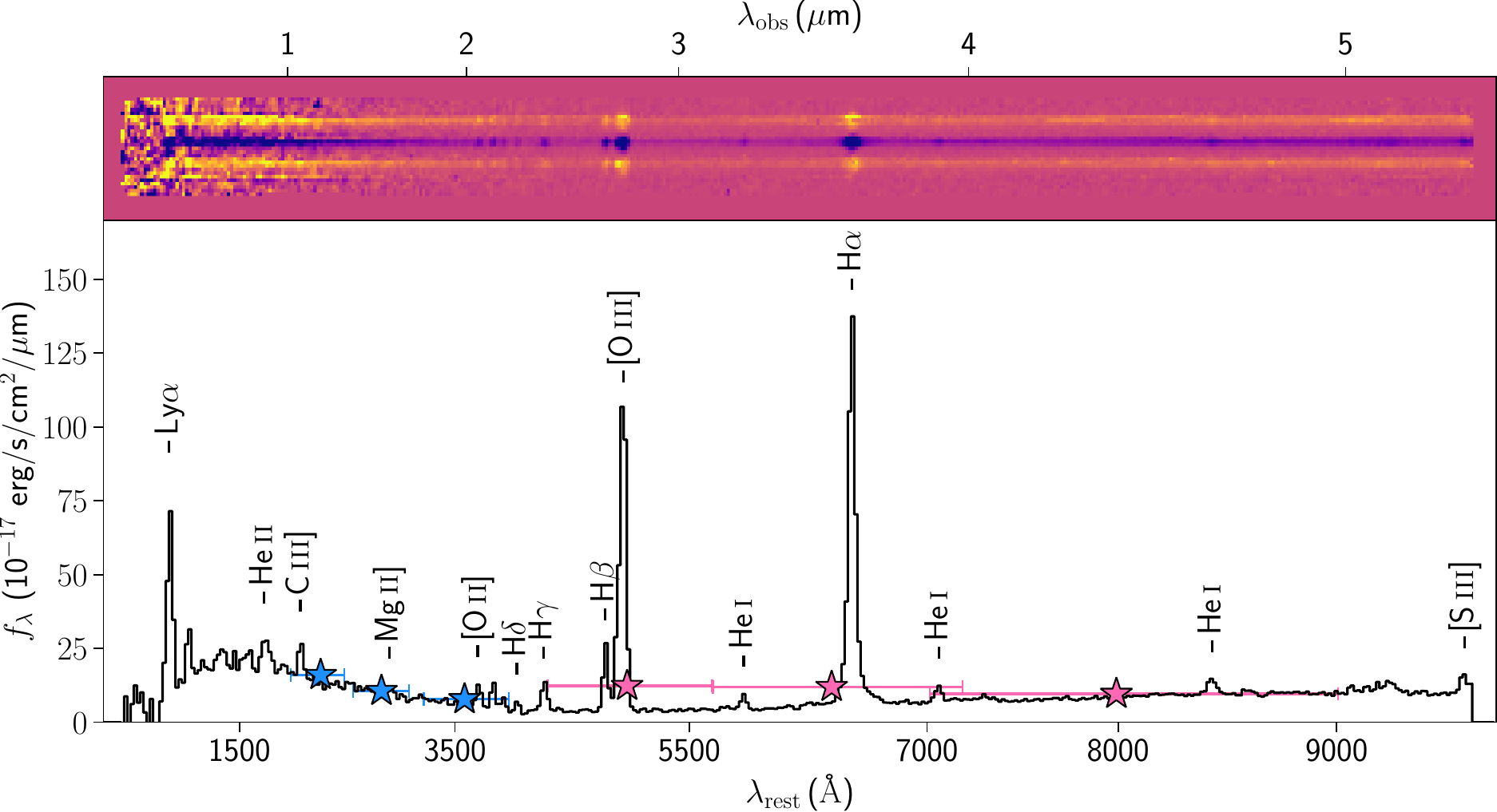}
    \caption[JWST observations of \lrd]{\textit{Top:} 2D NIRSpec PRISM spectrum of \lrd showing the bright \ha, \hb, and \oiii emission. \textit{Bottom:} 1D spectrum (corrected for Galactic extinction) with several identified lines labelled. The observed wavelengths are shown in $\upmu$m on the top axis, while the rest-frame wavelengths are plotted in \AA\ on the bottom axis. The NIRCam photometric fluxes in a 0.5\,arcsec diameter aperture in the UV filters F115W, F150W, and F200W are shown as blue stars with the NIRCam bands shown as errorbars. Optical/NIR filters F277W, F356W, and F444W are similarly shown as pink stars with errorbars.}
    \label{fig:filt_spec}
\end{figure*}

\lrd (RA=101.933406, Dec=70.198268) was observed in the MACSJ0647.7+7015 \citep{Ebeling2010TheSurvey} galaxy cluster lensing field as part of the \emph{JWST} Cycle 1 General Observers (GO) program (ID: GO-1433; PI: Dan Coe), including JWST/NIRCam imaging and high SNR JWST/NIRSpec prism spectroscopy. The NIRCam images in filters F115W, F150W, F200W, F277W, F356W, and F444W, and the NIRSpec PRISM/CLEAR MSA spectrum were obtained from the DAWN JWST Archive\footnote{https://dawn-cph.github.io/dja/index.html} (DJA; NIRSpec catalog ID 1045). The spectrum was extracted from the telescope exposures using \texttt{MsaExp v.0.6.7} \citep{Brammer2023Msaexp:Tools}, with standard wavelength, flat-field, and photometric calibrations. The reduction procedure is described in detail in \citet{Heintz2023Extremez=9-11}.

Fig.~\ref{fig:rgb_slit} shows an RGB image of the object, with NIRSpec slits overlaid. The 2D and 1D spectra \citep[optimally extracted following][]{Horne1986ANSPECTROSCOPY.} are shown in Fig.~\ref{fig:filt_spec}. The photometry from the six NIRCam filters in a 0.5\,arcsec aperture is overplotted on the 1D spectrum. The NIRSpec PRISM resolution varies across the spectrum with a minimum of R$\sim$30 at 1.2$\upmu$m, rising to a maximum of R$\sim$320 towards the NIR end \citep{Jakobsen2022TheCapabilities}. These resolution values are for uniform illumination of the slit and are higher for a compact source (see Sec.~\ref{sec:spectroscopy}). We correct the 1D spectrum for Galactic extinction using the \texttt{PlanckGNILC} map from the \texttt{dustmaps} library \citep{Green2018dustmaps:dust}. As there is good agreement between the spectrum and photometric fluxes, no slit-loss correction was done. Additionally, since \lrd is located at the outskirts of the lensing field, and its magnification is expected to be small, no magnification correction was applied.

\subsection{Morphology}
We model the galaxy's morphology in the six NIRCam filters using \texttt{GALFITM} \citep{Bamford2011MeasuringSurveys,Hauler2013MegamorphSurveys,Vika2013MegaMorph-MultiwavelengthFar}, a modified version of \texttt{GALFIT} 3.02 \citep{Peng2002DetailedImages,Peng2010DetailedModels} that can simultaneously fit multiple bands. Since \texttt{GALFITM} requires pixel matched images, we model the 20\,mas short wavelength UV bands (F115W, F150W, F200W) separately from the 40\,mas long wavelength optical/NIR bands (F277W, F356W, F444W). Each set of images are fit with three models: a point source, a S\'ersic profile, and a S\'ersic profile with a central point source. Point spread function (PSF) models for each band are generated with \texttt{WebbPSF}\citep{Perrin2012SimulatingWebbPSF,Perrin2014UpdatedWebbPSF}. Sigma images were derived from the weight maps provided by DJA. Object masks for background sources are created by dilating the DJA segmentation map by three pixels. Initial model parameters and sky background estimates are drawn from the DJA photometric catalogue for the MACS\,J0647 field.

\subsection{Spectroscopy}
\label{sec:spectroscopy}
Our spectroscopic model should account for the following features observed in LRDs: broad components in \ha and possibly \hb, narrow line emission, a blue continuum on the UV side, and a red continuum on the optical/NIR side.

We treat the UV and optical/NIR continuum as having different origins, modelled by independent components, and simultaneously fit emission lines with broad components added where necessary. In order to have control over the redshifts and velocity widths of individual components of each line, and properties of the two continuum components, while also accounting for the instrumental resolution, we use a custom fitting algorithm based on \texttt{lmfit} \citep{Newville2014LMFIT:Python}.

To fit the continuum, we use two power laws with distinct power law slopes extinguished by the Small Magellanic Cloud (SMC) dust extinction law \citep{Gordon2003ACurves}. We choose SMC extinction because with a power law model, the observed continuum requires a steeper attenuation than the Calzetti curve, for example, can provide \citep{Calzetti2000TheGalaxies}. In other words, the data are not consistent with a Calzetti-attenuated power law. However, a steepened Calzetti attenuation curve \citep[as in][]{Salim2018DustAnalogs} provides an equally good fit compared to the SMC extinction. The power law is of the form $f(\lambda) \propto \lambda^{-\beta}$, with $\beta$ constrained to be between 0 and -3 \citep{Bouwens2016MASS, Bouwens2023EvolutionHUDF/XDF}. Even with a fixed slope dust extinction law such as the SMC, there is a strong degeneracy between the intrinsic power law slope and the dust reddening, and therefore naturally a systematic uncertainty in the inferred extinction and/or the intrinsic slopes.

To fit emission lines, we identify a list of typical strong lines in star-forming galaxies in the wavelength region of our spectrum ($\sim$1000--9700 \AA; restframe). Each of these lines we model with a single Gaussian, with freely variable normalisation. We include an additional broad line for \ha (see Sec.~\ref{sec:widths}). For close doublet lines, we tie the Gaussian heights based on the intrinsic line ratios; for instance, we set the \oiii 5007 to 4959\,\AA\ ratio to be 2.98 \citep{Storey2000TheoreticalCases}. 
All the narrow line redshifts and widths are tied together. The Gaussian width is constrained to be above 100\,km\,s$^{-1}$ for narrow, and above 1000\,km\,s$^{-1}$ for broad lines (an arbitrary large value of 10\,000\,km\,s$^{-1}$ is set as the maximum for both). To obtain the Gaussian width of resolution-broadened lines, we convolve the input velocity widths with the \emph{JWST} prism resolution from \citet{Jakobsen2022TheCapabilities}. We scale the resolution by a constant factor of 1.3 to account for the improved resolution for compact sources in the NIRSpec slit \citep[see][]{deGraaff2023IonisedSpectroscopy, Greene2023UNCOVERz5}.

We estimate fit uncertainties by running an MCMC chain using the \texttt{emcee} fitting module in \texttt{lmfit}, with 
initial values set to the minimised result from our \texttt{lmfit} fit above. Given the large number of free parameters (37 for the full fit), we use 100 walkers, 10\,000 steps, and a burn-in phase of 100 steps to ensure that the chain is long enough to sample a sufficient portion of the parameter space.

\subsubsection{Analysis of line widths}
\label{sec:widths}
We conduct an analysis of emission line widths by fitting Gaussians of varying widths to individual lines. Given the low and uncertain resolution of our spectrum, we are unable to distinguish line widths below $\sim$1000\,km\,s$^{-1}$. Even in the highest resolution region of our spectrum towards the NIR end, we are limited to widths above $\sim$700 km/s. Indeed, it has been seen that the width of emission lines from point-like sources in the low resolution NIRSpec spectrum can vary by $\sim$1 wavelength bin \citep{Christensen2023MetalJWST/NIRSpec}. Hence, we cannot realistically distinguish between lines produced by star-formation ($\lesssim350\,$km\,s$^{-1}$) and AGN narrow line region (NLR) emission ($\lesssim1000\,$km\,s$^{-1}$). Nonetheless, we can measure broadness $\gtrsim1000\,$km\,s$^{-1}$, such as from an AGN broad line region (BLR). \ha shows a clear signature of broad emission. 
We try fitting a broad component to the \hb line as well, but find no evidence of broadness, and therefore do not include broad components for the other Balmer lines in the final fit (see Sec.~\ref{sec:alt_mods}).

\section{Results}
\label{sec:results}

\subsection{Morphology}
\label{sec:bestfit_morph}
\begin{figure*}
    \centering
    \begin{subfigure}[b]{\textwidth}
        \centering
        \includegraphics[width=\textwidth]{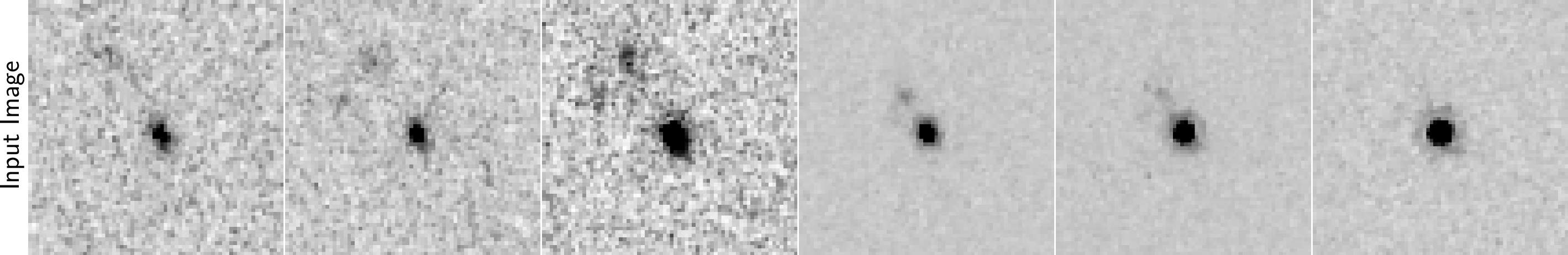}
    \end{subfigure}
    \medskip
    \begin{subfigure}[b]{\textwidth}
        \includegraphics[width=\textwidth]{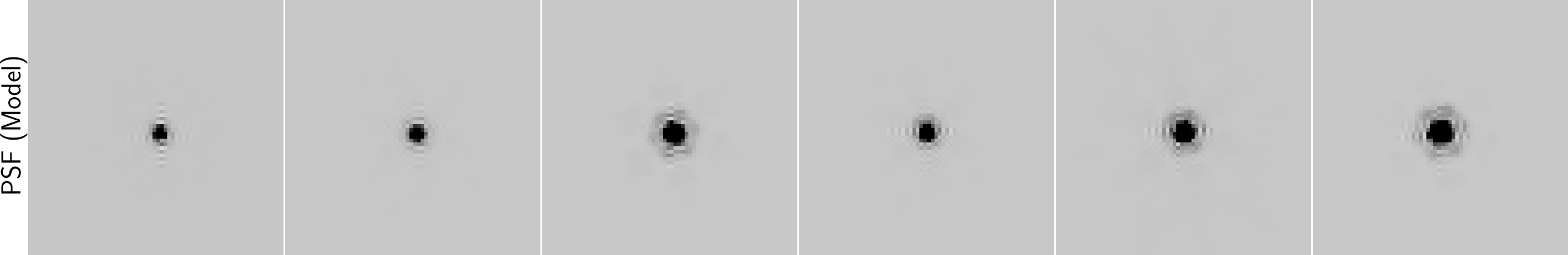}
    \end{subfigure}
    \medskip
    \begin{subfigure}[b]{\textwidth}
        \includegraphics[width=\textwidth]{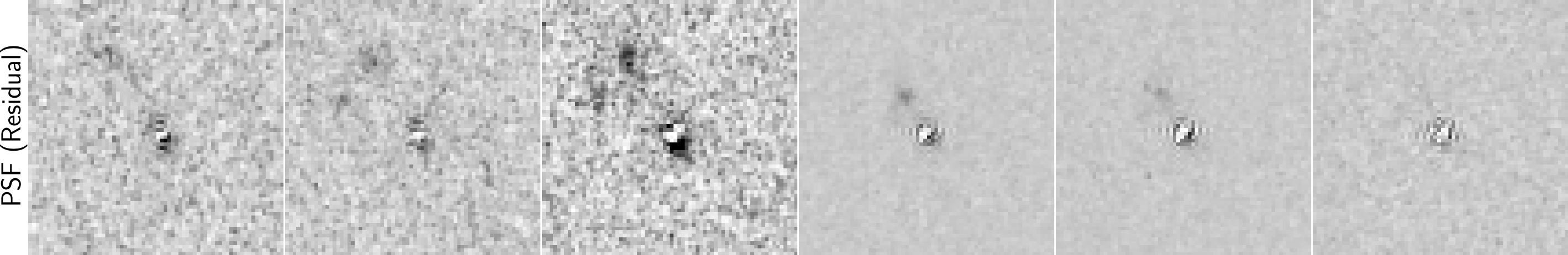}
    \end{subfigure}
    \medskip
    \begin{subfigure}[b]{\textwidth}
        \includegraphics[width=\textwidth]{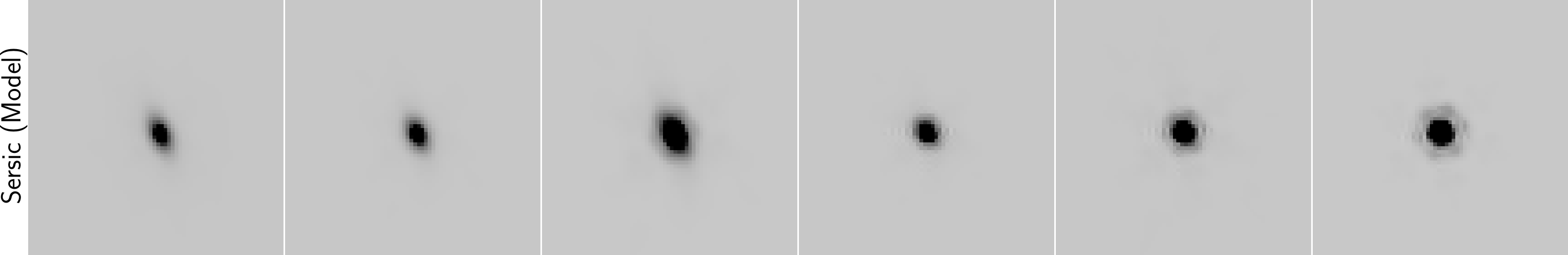}
    \end{subfigure}
    \medskip
    \begin{subfigure}[b]{\textwidth}
        \includegraphics[width=\textwidth]{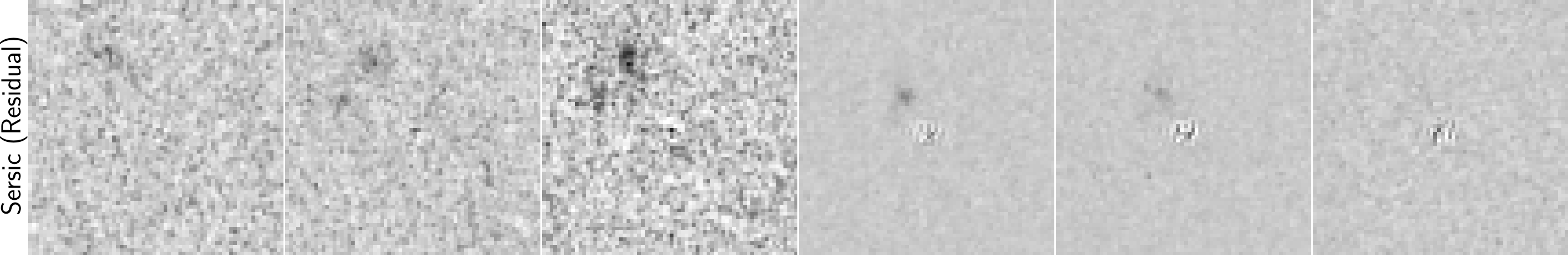}
    \end{subfigure}
    \medskip
    \begin{subfigure}[b]{\textwidth}
        \includegraphics[width=\textwidth]{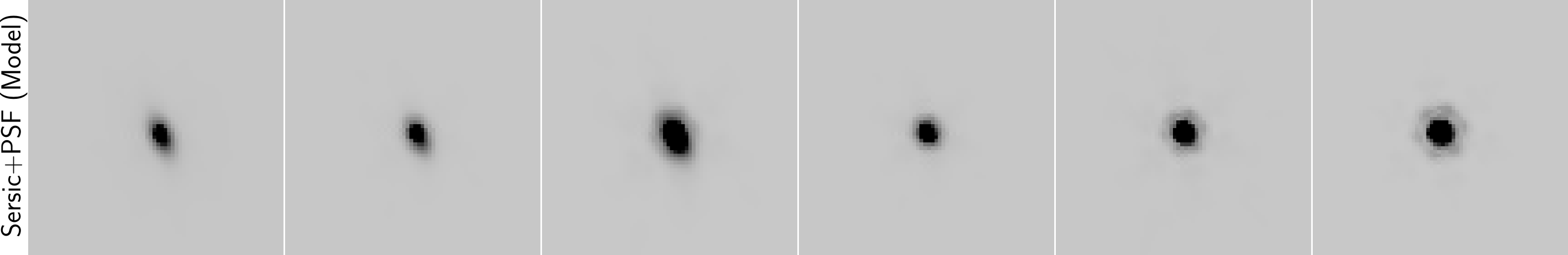}
    \end{subfigure}
    \medskip
    \begin{subfigure}[b]{\textwidth}
        \includegraphics[width=\textwidth]{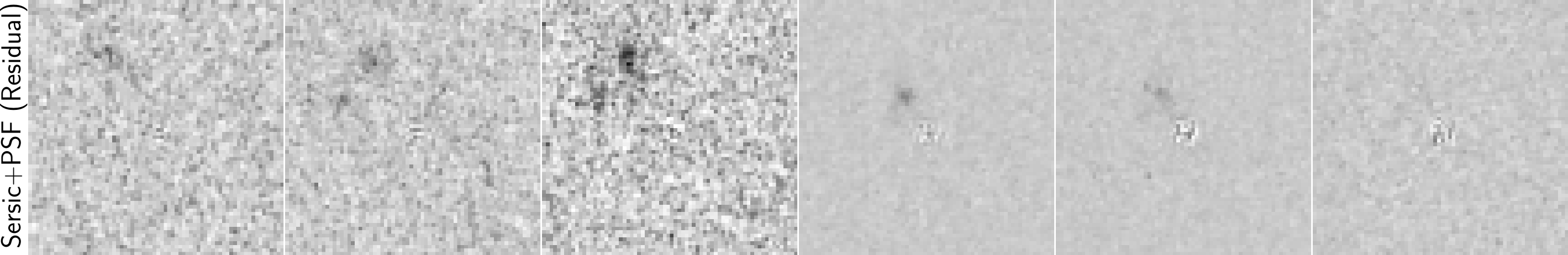}
    \end{subfigure}
    \caption[\lrd morphology fitting]{GALFIT models and residuals in the six NIRCam bands using PSF, S\'ersic, and combined S\'ersic+PSF models. The PSF fit shows clear residuals, especially in the UV bands. The fit improves considerably with the addition of a S\'ersic component. The best fit is obtained when both S\'ersic and PSF components are used, with $\Delta$BIC indicating a significant improvement in all bands (Sec.~\ref{sec:bestfit_morph}).}
    \label{fig:bestfit_morph}
\end{figure*}

We show our best-fit GALFIT model and residual in Fig.~\ref{fig:bestfit_morph}. The source appears to consist of a central component with extended irregular matter around it, visible especially in the UV bands. The best fit is obtained with a S\'ersic profile and a central point source, indicating the presence of a central compact core within an extended structure: an AGN within a host galaxy, for instance. We calculate the Bayesian information criterion (BIC) for these fits, and find $\Delta$BIC to be 2204 between the PSF and S\'ersic, and 2290 between PSF and S\'ersic+PSF fit for the UV bands. Similarly, $\Delta$BIC $\sim$ 7424 between the PSF and S\'ersic, and 8992 between PSF and S\'ersic+PSF fit for the optical/NIR bands. This indicates that the data fits better with the inclusion of both a PSF and a S\'ersic model. However, the size of the source (effective radius, $r_e \lesssim 0.17$--0.27\,kpc) is smaller than that of the PSF in the long-wavelength filters, implying that it is unresolved in the optical/NIR. The radius in the UV filters F115W, F150W, and F200W is $0.45\pm0.06$, $0.44\pm0.05$, and $0.38\pm0.03$\,kpc respectively (with GALFIT uncertainties re-scaled following Allen et\ al. (in prep.)). The change in radius may be an intrinsic feature of the morphology where the central AGN dominates towards the NIR, while the host galaxy contributes more to the continuum in the UV.

\subsection{Spectrum}
\label{sec:bestfit_spec}
\begin{figure*}
    \centering
    \includegraphics[width=\textwidth]{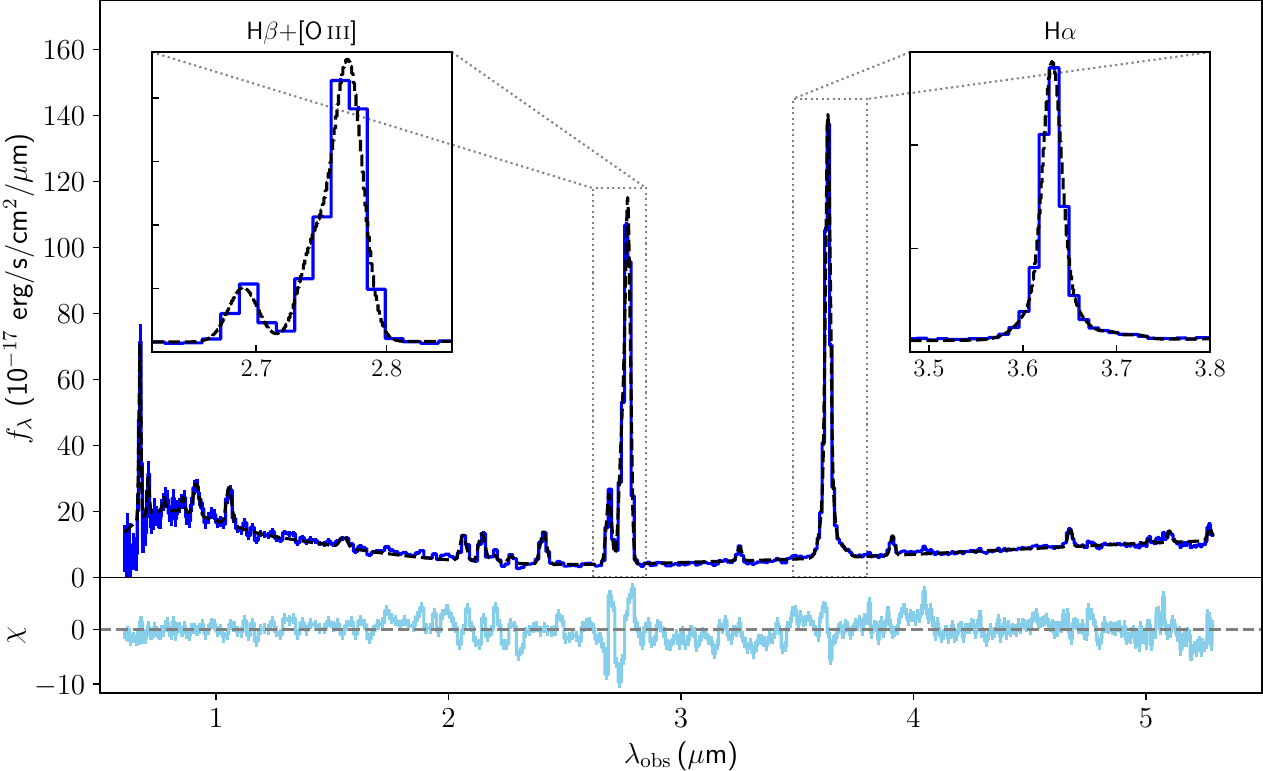}
    \caption[Best-fit to \lrd spectrum]{The observed spectrum (blue step curve) with the best-fit model (black dashed curve; plotted with a finer sampling than the observed spectrum), and zoom-in cutouts of \hb+\oiii and \ha regions. The residuals are shown in the bottom panel.}
    \label{fig:bestfit_spec}
\end{figure*}

\begin{figure}
    \includegraphics[width=\columnwidth]{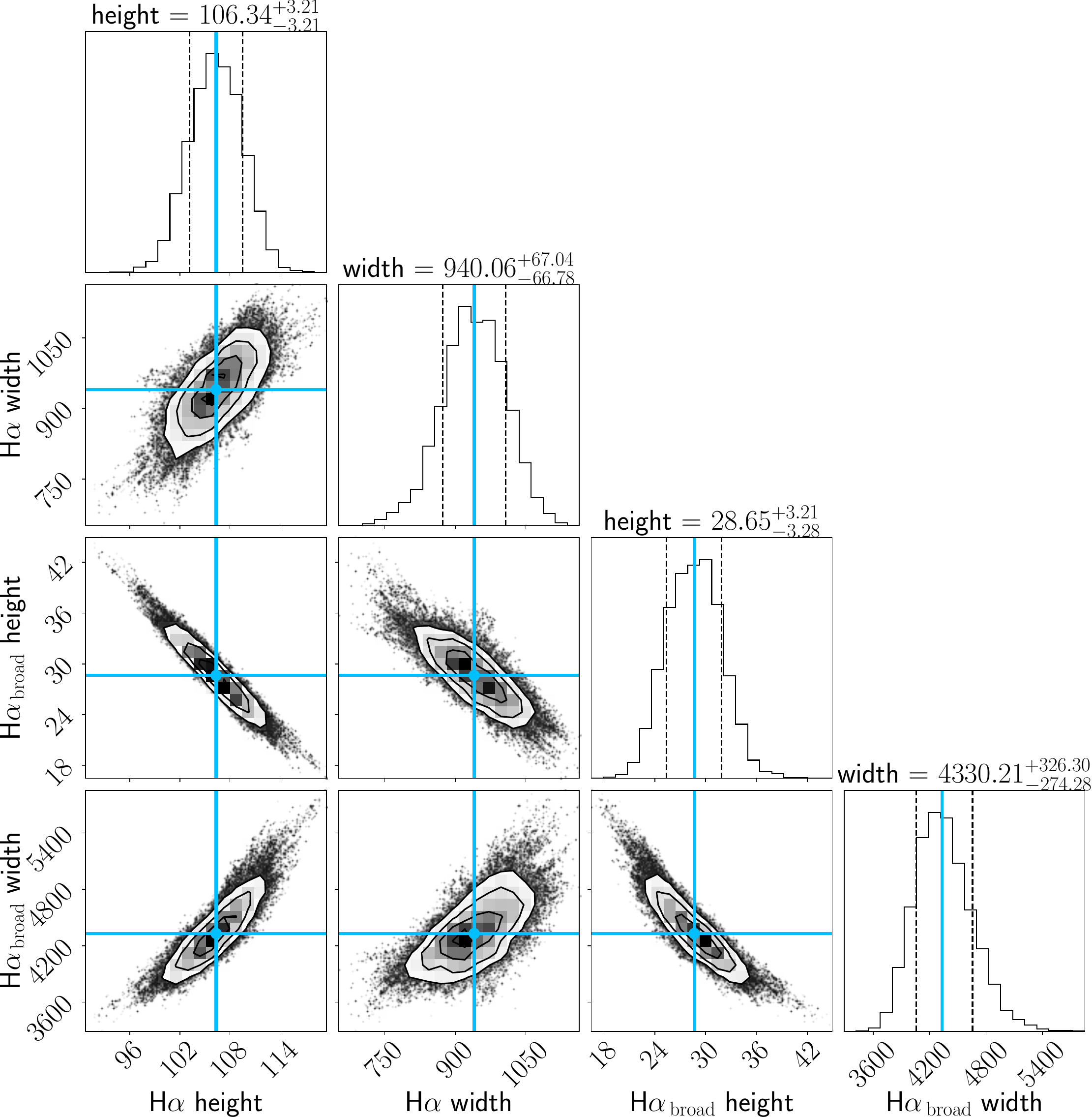}
    \caption[MCMC corner plot for \ha]{Corner plot showing the MCMC results for heights and widths of the narrow and broad \ha components. Height is given in the same units as Fig.~\ref{fig:bestfit_spec}, 10$^{-17}$\,erg\,s$^{-1}$\,cm$^{-2}\,\upmu$m$^{-1}$, and velocity width is given in km\,s$^{-1}$.}
    \label{fig:ha_corner}
\end{figure}

\begin{figure}
       \includegraphics[width=\columnwidth]{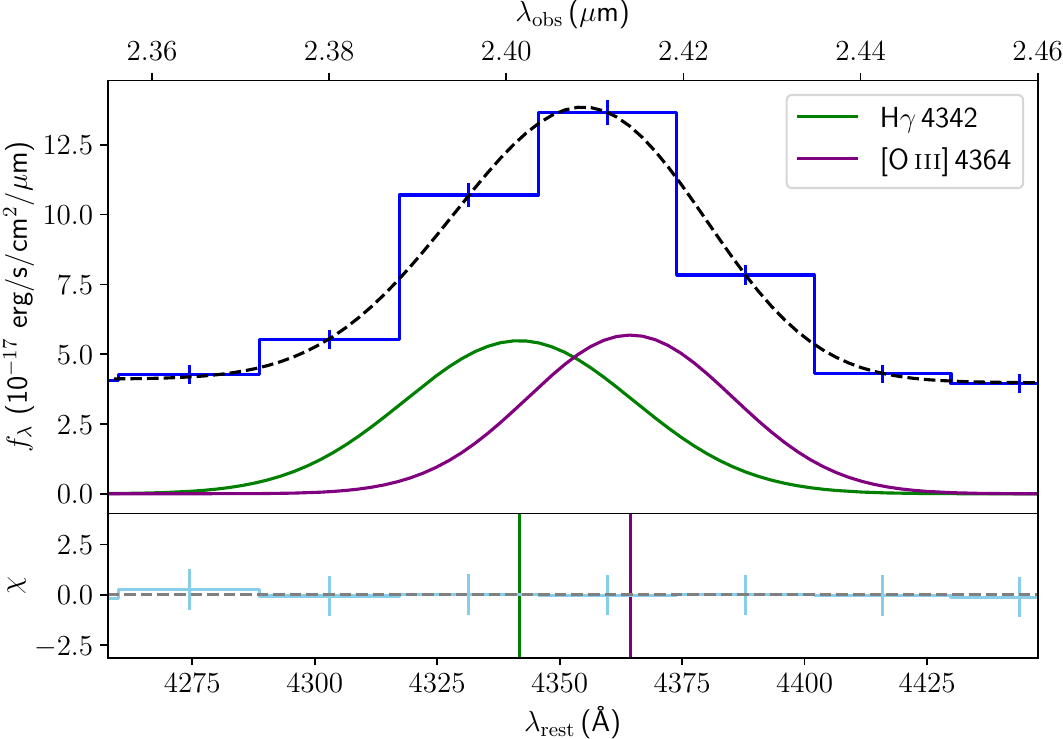}
       \medskip
        \includegraphics[width=\columnwidth]{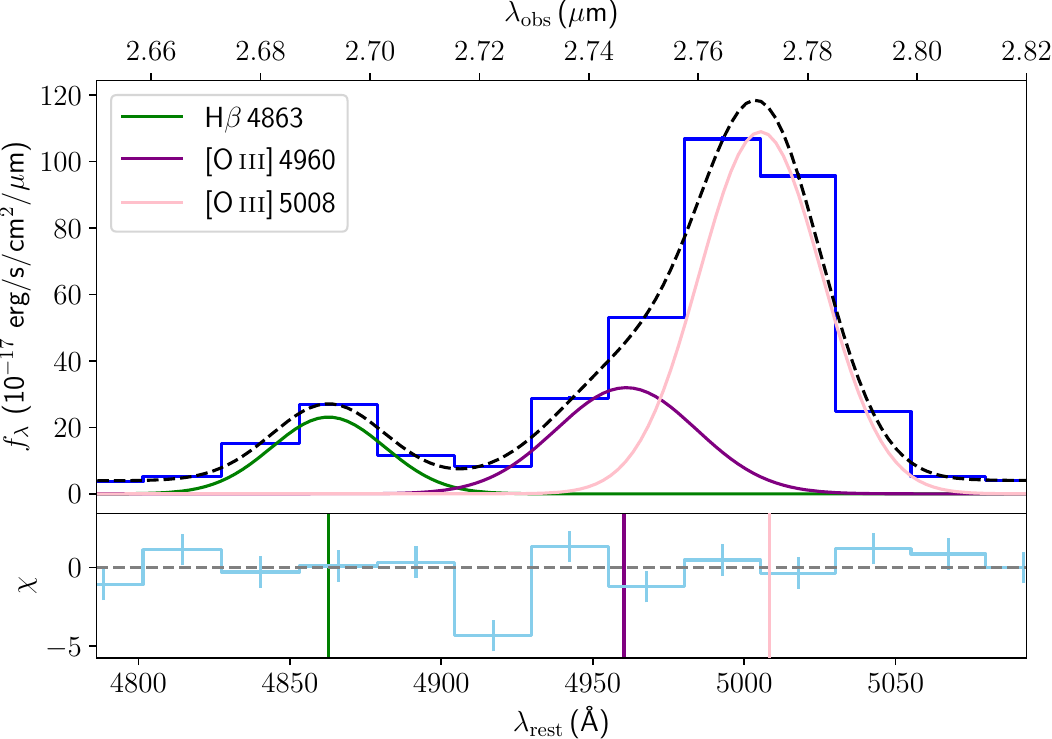}
        \medskip
        \includegraphics[width=\columnwidth]{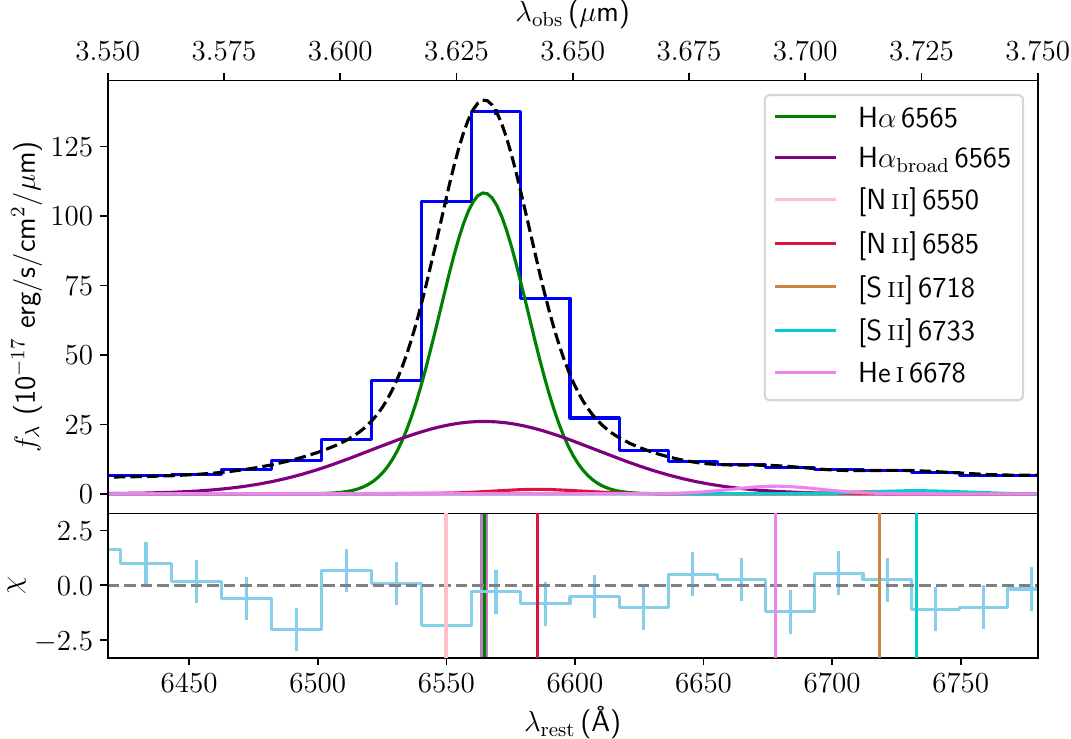}
    \caption[Best-fits to individual lines]{Best-fits to the \oiii and Balmer line complexes. The individual lines are indicated with coloured Gaussian curves, and their centre wavelengths are marked in the same colour in the residual panel below. The observed spectrum, best-fit, and residual curves follow the same colour scheme as Fig.~\ref{fig:bestfit_spec}.}
    \label{fig:individual_fits}
\end{figure}

The best fit to the spectrum is shown in Fig.~\ref{fig:bestfit_spec}. We find no appreciable velocity offset between the broad and narrow \ha lines when the \ha complex is fit separately, so we tie the narrow and broad redshifts together for the full fit. The MCMC uncertainties on broad and narrow \ha line heights and widths are shown in Fig.~\ref{fig:ha_corner}. The fit is not able to obtain a constraint on the \nii doublet (6549 and 6585\,\AA) flux in the \ha complex. Setting a fixed ratio of \nii / \ha = 1/100 \citep[which is the expected ratio for a metallicity of $\sim$0.1 solar;][]{Maiolino2019DeGalaxies} returns a poor residual, as the fit prefers a much lower ratio. Since we do not have the SNR to probe fluxes at this level, we do not include the \nii doublet in the final fit. Instead, to place constraints on \nii flux, we include it in the individual fit of the \ha complex (shown in the last panel of Fig.~\ref{fig:individual_fits}), and report the 1$\sigma$ upper limit from the corresponding MCMC run.

The redshift obtained for our best-fit model is $4.5321\pm0.0001$. This uncertainty includes only the statistical estimate from the MCMC. Altering model parameters results in redshifts that vary by 0.005 around the nominal value of 4.532, a consequence of the uncertainty in linewidths (see Sec.~\ref{sec:widths}). 
Moreover, a significant fraction of the residuals in our full spectrum fit are a result of the uncertainty on the dispersion function for the spectrograph. Hence, we fit several line complexes individually to obtain accurate line fluxes, without being influenced by variations in the continuum and instrument dispersion. We show these fits in Fig.~\ref{fig:individual_fits}. The best-fit results from all the above are summarised in Table~\ref{tab:bestfit_tab}.

\begin{table}
\caption[Best-fit \lrd properties]{Line wavelengths and fluxes from our best-fit model. Fluxes are given in units of $10^{-19}$\,erg\,s$^{-1}$\,cm$^{-2}$. \hei and broad \ha line fluxes are corrected by $A_V=5.7\pm0.2$ (indicated with a $\dagger$ symbol) from the optical-NIR power-law fit, while the rest of the lines are corrected by $A_V=1.1\pm0.2$ (indicated with a $\star$ symbol) from the Balmer decrement. The SMC attenuation correction is quite large at the \lya wavelength.}
\label{tab:bestfit_tab}
\begin{tabular}{@{}lccc@{}}
\hline\hline
Line & Wavelength & Flux & Flux\\
 & (\AA) & (uncorrected) & (corrected)\\
\hline
\lya$^\star$ & 1215.40 & 81$\pm1$ & 6$^{+12}_{-4} \times 10^{4}$ \\
{}\civ$^\star$ & 1549.48 & 17$\pm2$ & 2$^{+2}_{-1} \times 10^{3}$ \\
{}\heii$^\star$ & 1640.40 & 21$\pm3$ & 1$^{+2}_{-1} \times 10^{3}$ \\
{}\ciii$^\star$ & 1907.71 & 31$\pm4$ & 1$^{+1}_{-1} \times 10^{3}$ \\
{}\mgii$^\star$ & 2799.12 & 10$\pm4$ & 88$^{+58}_{-40}$ \\
{}\oii$^\star$ & 3728.50 & 25$\pm4$ & 127$^{+50}_{-36}$ \\
{}\neiii$^\star$ & 3869.87 & 26$\pm5$ & 128$^{+50}_{-36}$ \\
{}\neiii$^\star$ & 3968.59 & 9$\pm2$ & 40$^{+15}_{-11}$ \\
{}\hd$^\star$ & 4102.94 & 6$\pm3$ & 25$^{+19}_{-13}$ \\
{}\hg$^\star$ & 4341.73 & 20$\pm5$ & 79$^{+32}_{-24}$ \\
{}\oiii$^\star$ & 4364.44 & 21$^{+5}_{-6}$ & 80$^{+32}_{-26}$ \\
{}\hb$^\star$ & 4862.74 & 60$\pm4$ & 196$^{+31}_{-27}$ \\
{}\oiii$^\star$ & 4960.30 & 103$\pm1$ & 325$^{+68}_{-56}$ \\
{}\oiii$^\star$ & 5008.24 & 306$\pm4$ & 1$\pm0.2 \times 10^{3}$ \\
{}\hei$^\dagger$ & 5877.25 & 13$\pm3$ & 2$\pm0.4 \times 10^{3}$ \\
{}\ha (narrow)$^\star$ & 6564.70 & 256$\pm8$ & 562$^{+89}_{-78}$ \\
{}\sii$^\star$ & 6718.29 & 2$\pm0.3$ & 4$\pm1$ \\
{}\sii$^\star$ & 6732.67 & 4$\pm1$ & 8$\pm3$ \\
{}\hei$^\dagger$ & 7067.10 & 17$\pm2$ & 691$^{+126}_{-117}$ \\
{}\hei$^\dagger$ & 8446.70 & 16$\pm2$ & 247$^{+33}_{-31}$ \\
{}\ha (broad)$^\dagger$ & 6564.70 & 158$\pm6$ & 1$\pm0.1 \times 10^{4}$ \\
\hline
\end{tabular}
\end{table}

\begin{table}
\caption[SFRs and luminosities]{Dust-corrected luminosities and inferred star-formation rates. Luminosities are given in units of $10^{43}$\,erg\,s$^{-1}$. $L_{1500,\mathrm{UV}}$ is the intrinsic luminosity at 1500\,\AA\ from the UV continuum power-law fit, corrected for attenuation on the fit of $A_V=0.54\pm0.01$ (indicated with a $**$ symbol). $L_{5100,\mathrm{NIR}}$ is the luminosity at 5100\,\AA\ based on the optical/NIR fit, corrected by $A_V=5.7\pm0.2$ (indicated with a $\dagger$ symbol) from the fit. Dust correction for lines is the same as in Tab.~\ref{tab:bestfit_tab}.}
\label{tab:lum_sfr}
\begin{tabular}{@{}lcc@{}}
\hline\hline
Wavelength\,(\AA) & Luminosity & SFR (M$_\odot$\,yr$^{-1}$) \\
\hline
Ly$\alpha^\star$ & 123$^{+249}_{-82}$ & \\
\oii$^\star$ & 0.3$\pm0.1$ & $21\pm10$\\
\oiii$^\star$ & 3$^{+0.5}_{-0.4}$ & \\
H$\alpha$ (narrow)$^\star$ & 1$\pm0.2$ & $53\pm8$\\
H$\alpha$ (broad)$^\dagger$ & 22$\pm2$ & \\
$L_{1500,\mathrm{UV}}^{**}$ & 41$\pm1$ & $16\pm0.3$\\
$L_{5100,\mathrm{NIR}}^\dagger$ & 126$^{+11}_{-17}$ & \\
\hline
\end{tabular}
\end{table}

\subsection{Physical properties}
Using our best-fit model, we estimate the fluxes of narrow and broad emission lines. These values are all reported in Table~\ref{tab:bestfit_tab}. We estimate the Balmer decrement from the narrow \ha/\hb ratio following the prescription in \citet{Momcheva2013NEBULARSURVEY}, finding $A_V=1.1\pm0.2$, and use this to report attenuation-corrected narrow line fluxes. We also report line fluxes from \hei (see Sec.~\ref{sec:hei_lines}) and broad \ha, corrected for extinction using an $A_V$ of $5.7\pm0.2$, from the SMC extinction on the optical/NIR power-law continuum fit.

In Table~\ref{tab:lum_sfr}, we report the continuum luminosity at restframe 1500\,\AA\ ($L_{1500,\mathrm{UV}}$) and 5100\,\AA\ ($L_{5100,\mathrm{NIR}}$) from the UV and optical/NIR power law fits respectively. $L_\mathrm{1500,UV}$ continuum luminosity is corrected by an $A_V$ of $0.54\pm0.01$ from the attenuation on the UV power law fit. The uncertainty here is statistical, based only on the MCMC. In reality, $A_V$ is degenerate with the power-law slope, which in our fit approaches the constraint set at $-3$. $L_{5100,\mathrm{NIR}}$ is corrected by the extinction on the optical/NIR power law fit of $A_V = 5.7\pm0.2$, the same as for broad \ha line. The Balmer decrement of the broad \ha/\hb (using the upper limit on broad \hb) gives an $A_V>4.1$, which is consistent with the optical/NIR extinction. In other words, the optical/NIR extinction is sufficient to suppress a broad \hb component (see Sec.~\ref{sec:alt_mods}).

Using $L_{1500,\mathrm{UV}}$ luminosity from the UV fit, and correcting for dust attenuation using the $A_{V}$ from the fit, we obtain an SFR$_{\rm UV}$ of $16\pm1\,$M$_\odot$\,yr$^{-1}$, as indicated in Table~\ref{tab:lum_sfr}. Assuming the narrow \ha luminosity arises solely from star-formation, and correcting by the narrow line Balmer decrement attenuation of $A_V=1.1\pm0.2$, we obtain SFR$_{\mathrm{H\alpha}} = 53\pm8$\,M$_\odot$\,yr$^{-1}$, using the \citet{Kennicutt1998STARSEQUENCE} relation (and dividing by 1.8 to convert from Salpeter to Chabrier IMF). We also calculate SFR$_{\rm [OII]} = 21\pm10$\,M$_\odot$\,yr$^{-1}$ from the [O\,{\sc ii}] doublet lines at 3728\,AA, using eq. 3 of \citet{Kennicutt1998STARSEQUENCE}. Using Eq.~4 of \citet{Kewley2004Indicator} results in an SFR of $10\pm5$. The latter calibration seems reasonably appropriate for galaxies at $z=4.5$ \citep{Vanderhoof2022TheRelation}, though the scatter is quite large. The differences in inferred SFRs may be an indication that the narrow lines, especially \ha, are not exclusively powered by star-formation. We discuss this further below in Sec.~\ref{sec:bpt} and \ref{sec:hei_lines}.

The broad velocity width of the \ha line is $4300\pm300$\,km\,s$^{-1}$ (see Fig.~\ref{fig:ha_corner}). Using this velocity and the broad \ha luminosity, we estimate an SMBH mass of $8\pm1 \times 10^{8}$\,M$_\odot$ \citep{Greene2005EstimatingLine, Kocevski2023HiddenCEERS}. Using instead the continuum luminosity from the optical/NIR continuum at 5100\,\AA\ ($L_{5100,\mathrm{NIR}}$) results in a mass of $5\pm1 \times 10^{8}$\,M$_\odot$ \citep{Kaspi2000ReverberationNuclei, Kocevski2023HiddenCEERS}. These values agree within the uncertainty range for single-epoch measurements of SMBH mass \citep{Denney2009SystematicSpectra, Park2012TheEstimates, Campitiello2020EstimatingEpoch}. 
We estimate the bolometric luminosity to be $2.9\times10^{46}$\,erg\,s$^{-1}$ using the relation $L_\mathrm{bol} = 130 \times L_\mathrm{H\alpha, broad}$ \citep{Stern2012TypeProperties}.

Assuming the narrow line flux arises only from star-formation, the gas-phase metallicity (12+log(O/H)), following the R23, O32, and O2 formulations of \citet{Sanders2023DirectNoon}, is 7.3$\pm$0.1 (or 8.7$\pm$0.1 from the double-valued calibration), 7.9$\pm$0.3, and 7.8$\pm$0.3 respectively. This includes the intrinsic and fit uncertainties on the relation from \citet{Sanders2023DirectNoon}, and the uncertainties on line luminosities from our MCMC estimate. These values are consistent with the need to have a very low \nii flux to fit the \ha line as mentioned above, indicating a metallicity that may be significantly below 10\% of the solar value. A $\rm T_e$-based estimate using the \oiii\,4959, 5007\,\AA\ doublet along with \hb and \oiii\,4363\,\AA\ lines yields a similarly low metallicity of $7.0^{+0.2}_{-0.1}$ (\citealt{Perez-Montero2017IonizedMethod}; assuming the contribution from O$^{+}$ is small compared to that from O$^{2+}$). Although the density in  our case may be significantly higher than that assumed in this derivation, which would result in a slightly higher metallicity.

The ionisation parameter based on the sulphur and oxygen line flux ratios is relatively high. $\log{U} \sim -2.7$ from $\log$([\ion{S}{iii}]$\lambda\lambda$9069,9532/[\ion{S}{ii}]$\lambda\lambda$6717,6731), and $\log{U} \sim -2.4$ from $\log$([\ion{O}{iii}]$\lambda$5007/[\ion{O}{ii}]$\lambda\lambda$3726,3729) following \citet{Kewley2002Galaxies}. The reason for this discrepancy is unclear, but may be related to the low metallicity and ionisation to the 3+ ions.

Finally, we estimate the gas temperature using the ratio of \oiii\,4363\,\AA\ to \oiii\,4959+5007\,\AA\ doublet luminosity \citep{Nicholls2020EstimatingLimitations}. We obtain a value of $5\pm2 \times 10^{4}$\,K (agreeing with  our $\rm T_e$ estimate above), which is the typical temperature of ionised regions in AGN environments \citep[e.g.][]{Larson2023AQuasars}. However, a substantially higher density than typically assumed may allow for a somewhat lower, though still high, temperature.

\section{Diagnostics and models for the origin of the LRD phenomenon}
The origin of the emission in LRDs is perhaps the biggest open question currently. Based mostly on \ha lines with widths of about 1200--4000\,km/s, and on the very compact nature of these sources, AGN activity has been inferred in literature. Here we discuss the origin of the blue/UV and red/NIR continua, as well as the narrow and broad emission lines in \lrd, and consider several models that could reasonably fit the observed spectrum.

\subsection{Line emission}
\label{sec:alt_mods}
First, we check whether the broadness in the emission lines could arise from star-formation-related activity rather than an AGN. One possibility for line broadening is a merging system with multiple components at different redshifts \citep[e.g.][]{Maiolino2023JADES.Mighty}. To test this, we fit two sets of Gaussian lines each with common velocity widths $<1000\,$km\,s$^{-1}$ around the \ha line region. The two sets have different redshifts, and the line centres in each are together allowed to vary within $\pm 0.01 \upmu$m of their respective redshifts. The resulting model fails to fit the broad \ha feature. 

In some cases with star-formation in environments where the column density of atomic hydrogen is high, Raman scattering may broaden the Balmer lines to many thousands of km/s without strong Doppler broadening \citep[e.g.][]{Dopita2016THEREGIONS}. Such wings can be very broad in environments with very high column densities, but are likely to be relatively weak and have Lorentzian shapes \citep{Kokubo2023RayleighImplications}. Given the damped \lya absorbers with very high column densities observed in other high-redshift galaxies \citep{Heintz2023Extremez=9-11} and the compactness of LRDs, such extreme \hi columns might be present. We therefore test an instrumentally-broadened Lorentzian profile to the \ha line, but this does not fit the data well.

\begin{figure}
    \includegraphics[width=\columnwidth]{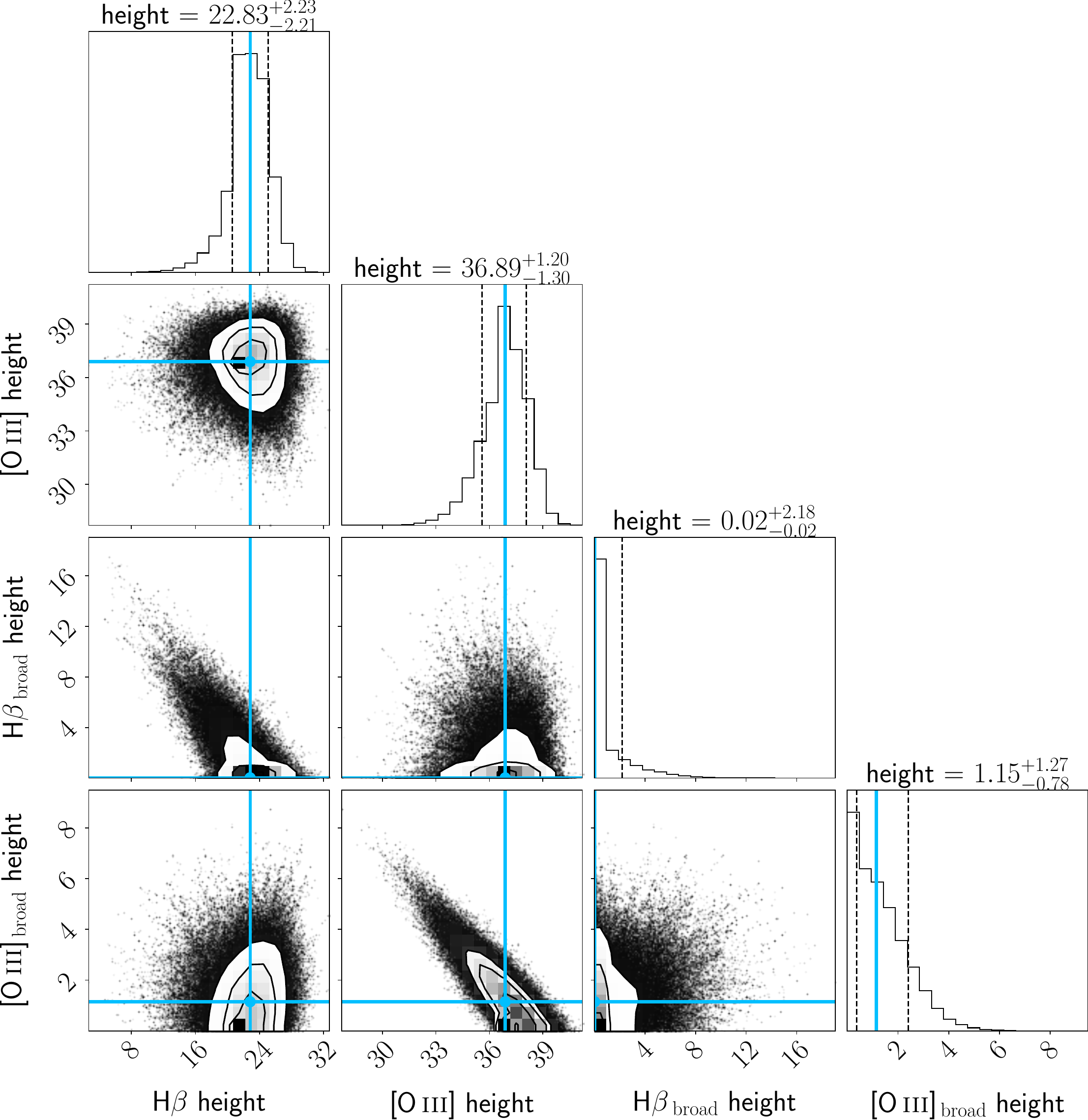}
    \caption[MCMC corner plot for \oiii and \hb fit]{Corner plot showing the MCMC results for broad component fits to \hb and \oiii doublet (4959 and 5007\,\AA) lines. Height is given in the same units as Fig.~\ref{fig:bestfit_spec}, 10$^{-17}$\,erg\,s$^{-1}$\,cm$^{-2}\,\upmu$m$^{-1}$.The height of the broad component is arbitrarily small for both \oiii and \hb.}
    \label{fig:no_oiii_broad}
\end{figure}
The \ha line is best-fit with an unresolved narrow component and a broad component over 4000km/s -- a velocity much higher than expected from a star-formation or even extreme supernova driven outflows \citep{Fabian2012ObservationalFeedback, Baldassare2016EMISSION, Davies2019KiloparsecSurvey}. We also find no significant offset between the narrow and broad \ha components, which would be expected for an outflow. Moreover, we do not find any evidence of a broad component in the \oiii doublet at the same width and strength (relative to the narrow component) as the broad \ha component. We show the results of our MCMC analysis in Fig.~\ref{fig:no_oiii_broad}, by which we exclude the possibility of a broad \oiii line. We therefore conclude that the most likely origin for the broad \ha line is an AGN BLR.

Curiously, we fail to find any significant broad component in \hb (Fig.~\ref{fig:no_oiii_broad}). This is not due to the relatively lower SNR of the \hb line compared to \ha. The \ha/\hb ratio using the broad \hb upper limit from the MCMC fit is $\gtrsim13$. This ratio translates to an $A_V$ of $\gtrsim4.1$ for an SMC extinction curve, consistent with the $A_V$ of 5.7$\pm$0.2 from the optical/NIR continuum in our best-fit model, suggesting a model whereby the broad lines and red/NIR continuum are extinguished by a similar dust column. In contrast, the narrow \ha/\hb ratio is only $\sim$4, which gives an $A_V$ of $1.1\pm0.2$. From our best-fit model, we obtain an $A_V$ of $0.54\pm0.01$ for the UV continuum (and a power-law slope approaching -3). In other words, the UV continuum and narrow emission lines have similar modest dust obscuration, much lower than the red/NIR continuum and broad emission lines. We conclude that there must be two distinct origins for the narrow and broad emission lines, consistent with the origins of the blue/UV and red/NIR continua respectively.

\subsection{Diagnostic diagrams}
\label{sec:bpt}
\begin{figure*}
    \centering
    \begin{subfigure}[b]{0.5\textwidth}
        \centering
        \includegraphics[width=\textwidth]{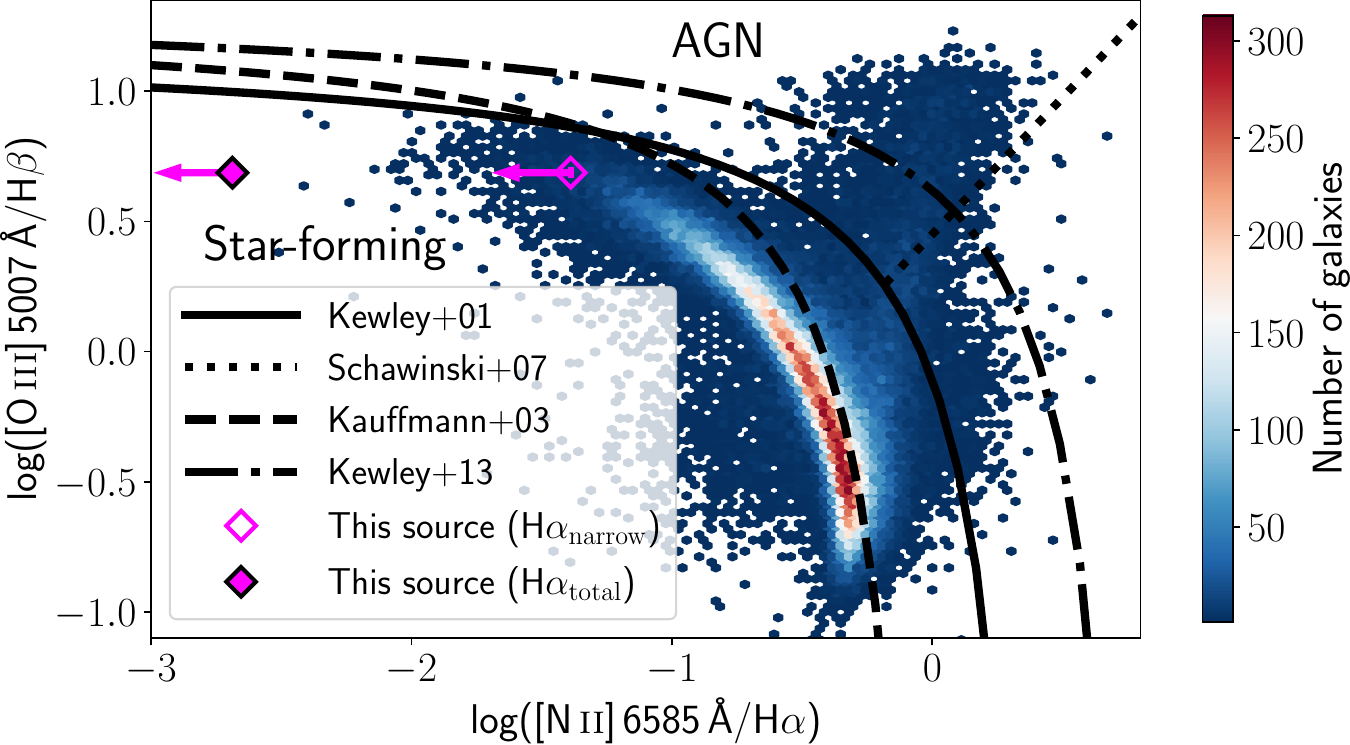}
    \end{subfigure}
    \hfill
    \begin{subfigure}[b]{0.47\textwidth}
         \centering
         \includegraphics[width=\textwidth]{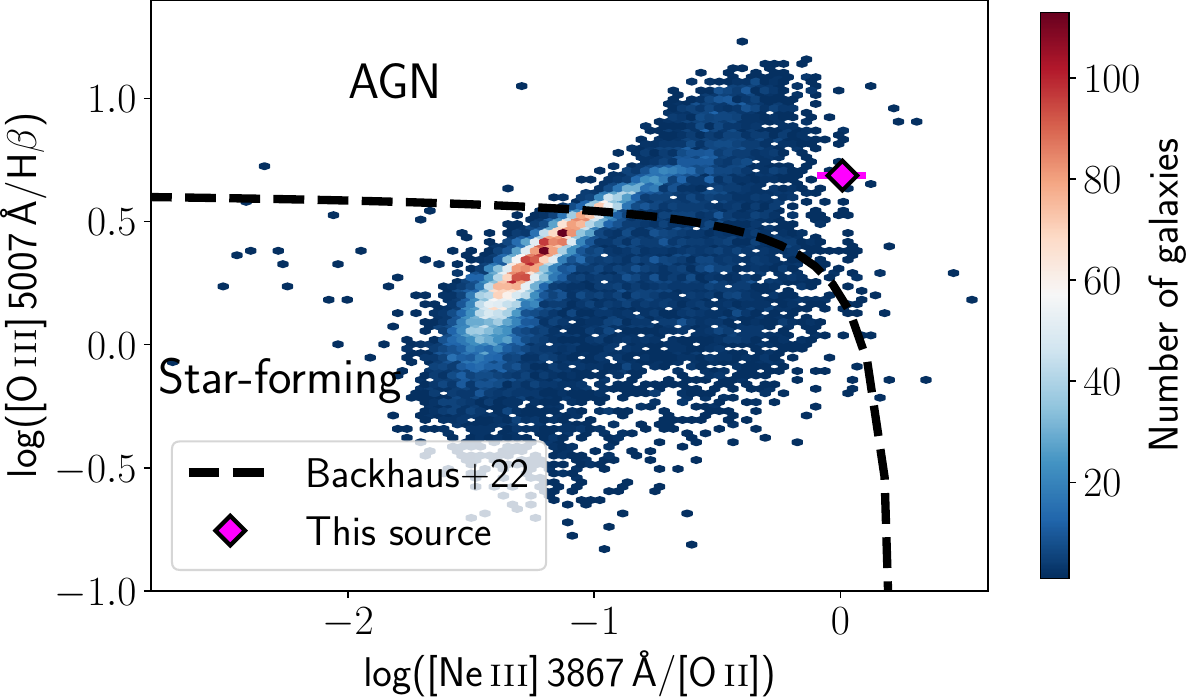}
    \end{subfigure}
    \caption[\lrd on diagnostic diagrams]{\textit{Left:} BPT diagram of log(\nii/\ha) flux vs log(\oiii/\hb) flux of SDSS galaxies. \nii flux is an upper limit of the 6585 \AA\ line, \oiii flux is the sum of the (4959, 5007 \AA) doublet, and \hei flux is for the 5876 \AA\ line. The log(\nii/\ha) upper limit for this source, using only the narrow \ha component flux and the total flux (sum of narrow and broad component fluxes), is shown by open and filled magenta diamonds respectively. The theoretical cutoff curves for AGN and star-forming galaxies at $z=0$ from \citet{Kewley2001TheoreticalGalaxies, Schawinski2007ObservationalGalaxies, Kauffmann2003TheAGN}, and the redshift-dependent curve from \citet{Kewley2013THEOBSERVATIONS} are shown as black solid, dotted, dashed, and dash-dot lines respectively. \textit{Right:} OHNO diagram of log([Ne\,{\sc iii}]/[O\,{\sc ii}]) vs log(\oiii/\hb) flux of SDSS galaxies. The curve separating AGN and star-formation from \citet{Backhaus2022CLEAR:Noon} is plotted as a black dashed line. The position of this source is again marked by a magenta diamond. [O\,{\sc ii}] luminosity is the sum of the (3726, 3729 \AA) doublet, modelled as a single Gaussian in our fit.}
    \label{fig:bpt}
\end{figure*}

The BPT diagram \citep{Baldwin1981ClassificationObjects., Kewley2013THEOBSERVATIONS} is a common diagnostic used to separate AGN and star-forming sources. As our best-fit model is unable to recover the \nii flux (Sec.~\ref{sec:bestfit_spec}), we can only obtain an upper limit on the \nii/\ha ratio. The BPT diagram (Fig.~\ref{fig:bpt}) suggests an extreme star-formation origin for the narrow lines in \lrd. However, this may not be definitive since several studies find that the BPT diagram mis-classifies LRDs due to their low metallicity \citep[e.g.][]{Harikane2023JWST/NIRSpecProperties, Ubler2023GA-NIFS:IFS}. The OHNO diagram has been proposed as an alternative \citep{Kocevski2023HiddenCEERS}. We plot both in Fig.~\ref{fig:bpt} and compare with $z\sim0$ galaxies from the Sloan Digital Sky Survey (SDSS).

The OHNO diagram also indicates a hard radiation field for the narrow lines in \lrd, but on the AGN side of the boundary for $z=0$ galaxies. This suggests that the narrow line emission in \lrd may be a combination of the star-formation in the galaxy and light from the AGN narrow-line region. Indeed, the SFR inferred from the narrow component of \ha alone is a factor of a few greater than the SFR inferred from the $L_\mathrm{1500,UV}$. While this could be due to the different timescales of these SF indicators, the fact that the metallicity-corrected [O\,{\sc ii}] SFR is also much lower than the \ha SFR, hints that AGN activity may contribute to the narrow line fluxes. 
Finally, the electron temperature we infer from the [O\,{\sc iii}] lines, as discussed in Sec.~\ref{sec:results}, is very high and may more likely be explained with a combination of a moderately high temperature and very high density, again, hinting at an AGN contribution to these lines.


\subsection{He emission}
\label{sec:hei_lines}
We identify several \hei emission lines across the spectrum. Measuring $A_V$ from \hei 5877 and 7065\,\AA\ line fluxes gives a value consistent with that obtained from our optical/NIR continuum fit, and broad line Balmer decrement. This points to an AGN, rather than star-formation, origin for these lines. However, as the SNR of these lines is low, we do not include broad Gaussian components to these lines in our model. Further, the ratios between these lines and \hei\,3889 and 8446\,\AA\ are not consistent with the same $A_V$, although the latter may be blended with other lines (\neiii\,3869\,\AA\ and \oi\,8446\,\AA), affecting our estimate. We also measure a flux ratio between \ciii\,1909\,AA \citep[obtained assuming \ciii\,1907/1909 = 1.53;][]{Maseda2017TheSurvey, Kewley2019UnderstandingLines} and \heii\,1640\,\AA\ of $0.6\pm0.1$, which implies a harder radiation field than can be produced via star-formation \citep{Kewley2019UnderstandingLines}.

\subsection{Origin of the continuum}
\subsubsection{Thermal emission from dust torus}
Motivated by the compact-dominated nature of the morphology in the long wavelength bands, we consider whether the rest optical/NIR continuum (i.e.\ the red spectral side) could be modelled as direct thermal dust emission from the inner edge of an AGN torus, i.e.\ a blackbody curve instead of a power law. We also add variable SMC extinction, just as we did with our power-law model. This model fits the optical/NIR continuum data well, with a dust temperature of T$_b \sim$ 2500\,K and no significant extinction required on the optical/NIR side. 
However, this temperature is significantly hotter than type~1 AGN dust tori, which are typically closer to 1400\,K \citep{Kishimoto2007TheReverberation, Honig2010TheAGN}, and it is even hotter than models of carbon grain thermal sublimation temperatures, which are around 2000\,K \citep{Kobayashi2009DustModel}. Moreover, if the broad-line region was obscured, we would expect significant dust-obscuration of the blackbody emission from the inner edge of the torus region, in contrast to the very low dust-obscuration obtained from our blackbody fit. It may be possible at certain viewing angles of the AGN for the blackbody emission from the inner dust torus to be directly transmitted to the observer without being screened by dust, but if this were the case, we would not expect to see such large obscuration of the broad lines. Hence, while the continuum fit is good, the high temperature and inconsistency of the dust extinction with the broad Balmer lines leads us to disfavour the possibility that the optical/NIR continuum arises from thermal emission from an AGN dust torus.

The UV $\beta$ slope and $A_V$ in our continuum fit are highly degenerate, but the fit favours $\beta$ slopes approaching $-3$ (where $F_\lambda \propto \lambda^{\beta}$). Such extreme blueness of the dust-corrected UV slope seems hard to explain, whether for an AGN \citep[e.g.][]{Hjorth2013OnDust} or for a young star-forming population \citep{Bouwens2016MASS}. However, fixing the power-law slope to $-2.7$ does yield an acceptable fit. Another curious aspect of the continuum, which is beyond the scope of this paper, is the rollover of the spectrum at the \lya break. We could not replicate this feature using dust attenuation, but suggest that it may indicate \lya damping in addition to dust attenuation \citep{Heintz2023Extremez=9-11}. Although the presence of the \lya line in emission could be difficult to explain in that case.



\subsubsection{Origin of the `V'-shape from an AGN core extinction curve?}
\begin{figure}
    \includegraphics[width=\columnwidth,clip=]{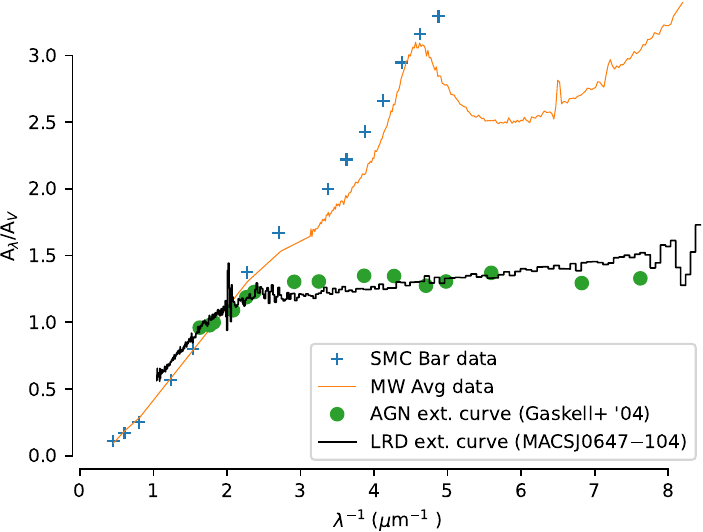}
    \caption{The extinction curve of \lrd assuming a single power-law intrinsic spectrum. The AGN core emission extinction data of \citet{Gaskell2004TheDistribution} are shown for comparison alongside the Milky Way average extinction from \citet{Gordon2009FUSESightlines} and the SMC bar extinction data from \citet{Gordon2003ACurves}. The close match to the \citet{Gaskell2004TheDistribution} curve data is suggestive, and indicates that 
    the observed V-shaped continuum of the little red dots could be produced by a single power-law continuum with peculiar AGN extinction.}
    \label{fig:extinction_curve}
\end{figure}

It is noteworthy that the location of the break in the continuum is around 4500\,\AA, close to the location of the flattening in the proposed AGN core extinction curve of \citet{Gaskell2004TheDistribution}. Is it possible that the `V'-shape observed in LRDs is simply the result of the flattening UV extinction curve of AGN? To test this hypothesis, we derive an observed extinction curve for \lrd by assuming the best fit red continuum $A_V$ to set the $V$-band normalisation of the intrinsic spectrum, and then assume a very blue intrinsic slope ($\beta=-2.7$). We then divide the observed spectrum by this single power-law and normalise by the $A_V$. This extinction curve is shown in Fig.~\ref{fig:extinction_curve} along with the data used by \citet{Gaskell2004TheDistribution} to derive the AGN continuum extinction curve. The match is surprisingly good (though it must be admitted that it is difficult to differentiate this type of extinction curve from one produced by a partial covering of a high-extinction sightline). We are not too concerned with an exact match to these data here, as other assumed intrinsic normalisations and slopes will change the shape of the extinction curve somewhat. In addition, some contribution from the host galaxy must be present. There is also a range in the AGN extinction curve properties derived for other AGN with slightly different break wavelengths and UV slopes \citep[e.g.][]{Gaskell2017TheSizes}. However, overall the match is similar, showing strong reddening up to about 4000\,\AA\ and then a much flatter extinction in the blue/UV. These flat UV extinction curves have been argued to be the product of the thermal sublimation of smaller grains by the AGN UV and X-ray emission. However, it is unclear why the AGN would destroy small grains deep into the torus, and in that case, the extinction curve would only be UV-flat in regions with relatively modest UV extinctions, i.e.\ with $A_{\rm UV}\lesssim 1$. For this reason, we would expect the overall extinction curve to be UV-flat only for a small subset of AGN viewed at specific angles and with modest extinctions. This cannot be the case for \lrd or most LRDs. Hence, in spite of the notable agreement between the V-shaped SED phenomenon in LRDs and a single power-law fit with AGN-core extinction, we do not currently favour this explanation.

\subsubsection{Other origins of the V-shape}
The apparently V-shaped SEDs of LRDs have been suggested to be due to an AGN observed in dust-obscured direct emission and low-extinction scattered emission \citep{Labbe2023UNCOVER:ALMA}, or obscured star-formation with the blue component due to direct AGN emission \citep{Kocevski2023HiddenCEERS}. Another possibility considered is a partial coverer, with a small fraction of low-extinction emission (A.~Goulding, priv. comm.). In all these cases, the blue light would be expected to be dominated by a point source, possibly compatible with the imaging data for this source. However, the Balmer decrement should also be small for the broad lines for all these hypotheses, since the dust obscuration should not affect the scattered broad lines very much. Since we do not detect broad \hb, none of these hypotheses seem compatible with the data in our case. A Balmer break was also proposed \citep{Labbe2023UNCOVER:ALMA} but seems unlikely to produce the observed features here.

\subsection{Are we observing the AGN core directly through a dust screen?}

One possible interpretation of the LRD phenomenon is a type~2 AGN in a star-forming host galaxy. While at lower redshift, most AGN are in relatively metal-rich hosts, as we go to high redshifts, the host galaxies will be more metal poor and therefore also more dust poor. It is not inconsistent with the existing data to argue that the LRDs may be type~2 AGN in low-metallicity hosts, viewed directly through their obscuring ``dusty torus''. At lower redshifts, the metallicity is closer to the solar value and the gas-to-dust ratio is close to Galactic. In a scenario where the total gas column density through the torus is above $\log{N_H/\mathrm{cm}^{-2}} = 23-24$, the typical $A_V$ will be $50-500$ magnitudes at Galactic gas-to-dust ratios, and these AGN would only be detected along their lowest extinction orientations at optical/NIR wavelengths.

However, at $z>4$, where the host galaxy metallicity may be 1--10\% of the solar value, the $A_V$ could be only a few magnitudes, since the dust-to-gas ratio is likely to be 20--1000 times lower than the Galactic value at these metallicities \citep{Konstantopoulou2023DustCorrigendum}. This would allow these AGN to be detected directly through their dust tori. In this scenario, we would then predict low metallicities to be typical of LRDs. In the very lowest metallicity cases, strong molecular hydrogen absorption bands would be expected in the UV, however in almost all cases, the Lyman-Werner bands will not be detectable due to the suppression of the UV flux at even very low metallicity and blueward of \lya by intergalactic \ion{H}{i} at high redshifts. 

Whether LRDs really are the direct detection of the elusive population of type~2 AGN at high-redshift could be tested based on number counts, X-ray limits, and spectra. The current lack of X-ray detections \citep{Matthee2023EIGER.JWST,Furtak2023AShadows} is already somewhat problematic for an AGN scenario, and seems to require lower mass black holes and super-Eddington accretion \citep{Greene2023UNCOVERz5,Fujimoto2022ADawn}. 

Photoelectric absorption through the torus could significantly suppress the X-ray emission even in the low metallicity scenario. For example, for a spectrum taken through gas clouds with say 2\% solar metallicity, which corresponds to a dust-to-gas ratio of $10^{-5}$ \citep{Konstantopoulou2023DustCorrigendum}, and assuming an $A_V=6$, would require an actual (not equivalent) hydrogen column density of $\sim10^{25}$\,cm$^{-2}$. Soft X-rays will still be significantly absorbed below about 1\,keV in the observed frame for systems at $z\sim5$. However, for these high-redshift systems, harder X-rays should still be detectable. In this hypothesis, the sources are unlikely to be Compton thick since the dust-to-gas ratios inferred would have to be well below expectations for the sort of metallicities these galaxies seem to have. 

\subsection{Comparison with other LRDs}
\begin{figure}
    \centering
    \includegraphics[width=0.5\textwidth]{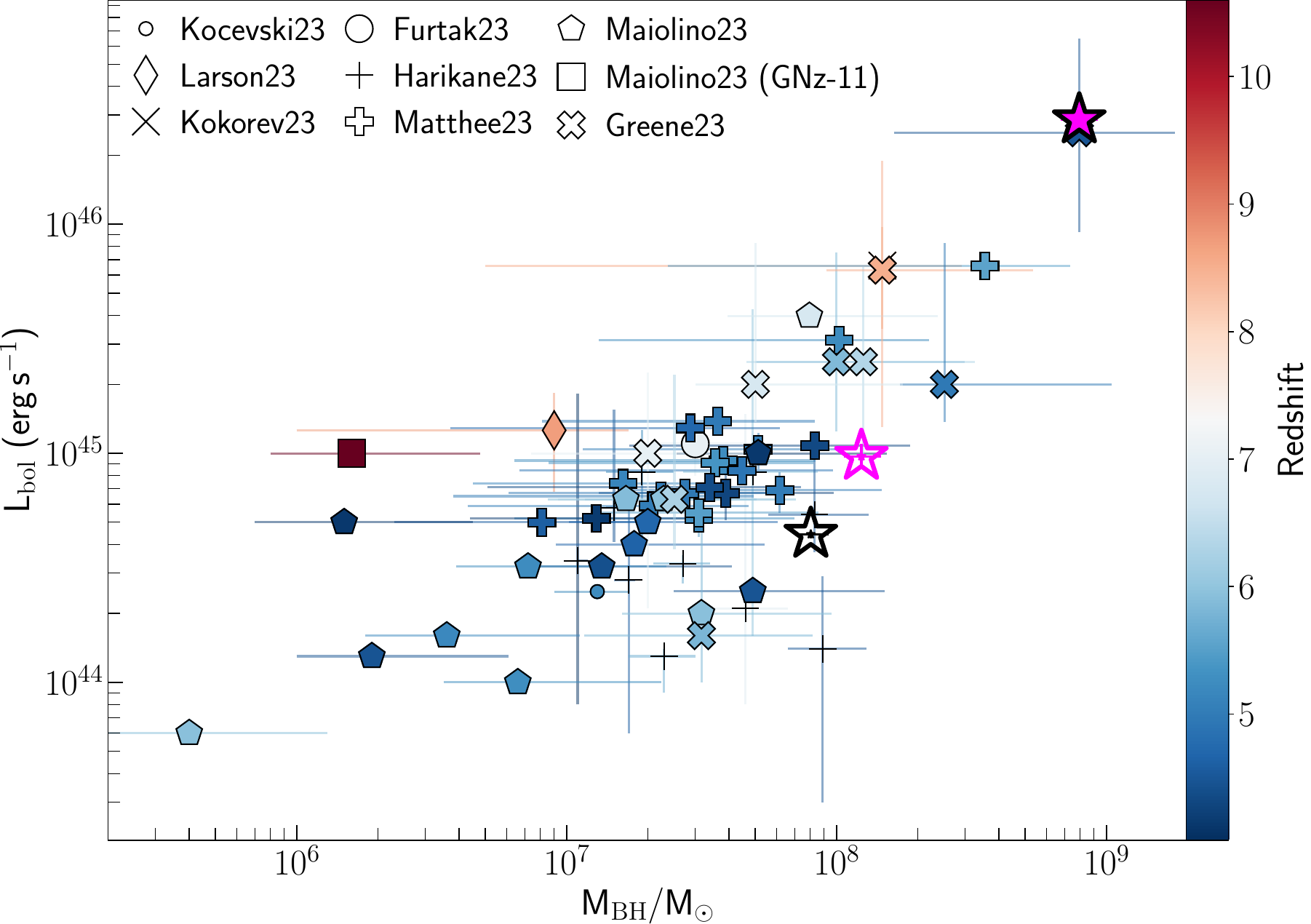}
    \caption[SMBH mass and Lbol]{SMBH mass and bolometric luminosity of \lrd in comparison with measurements for other LRDs from the literature. The symbols are as indicated in the legend, and the colour-scale represents redshift. The filled magenta star shows SMBH mass and $L_\mathrm{bol}$ estimates for our source, derived from broad \ha line using a dust correction by $A_V=5.7\pm0.2$ corresponding to the AGN extinction from our model. The empty magenta star shows these properties corrected by $A_V=1.1\pm0.2$ corresponding to Balmer attenuation. The empty black star shows measurements for our source with no dust correction applied.}
    \label{fig:Mbh_Lbol}
\end{figure}

In Fig.~\ref{fig:Mbh_Lbol}, we plot the bolometric luminosity against SMBH mass for \lrd and other LRDs from the literature. Some of these works have corrected their estimates for dust attenuation \citep[e.g.][]{Greene2023UNCOVERz5}, while others \citep[e.g.][]{Matthee2023EIGER.JWST} have not corrected for dust. For a fair comparison with such cases, we plot the measurements for our source with Balmer and AGN dust correction, and with no dust correction. As can be seen, if AGN dust correction is applied, \lrd is among the most extreme LRDs detected so far.

\subsection{Future avenues}

There are several directions to explore the LRD phenomenon and its connection to the obscured AGN population. One possibility is to conduct large \emph{JWST} surveys targeting this population, which can help set constraints on the AGN number counts in the early Universe. Preliminary efforts in this direction have already proven promising \citep[e.g.][]{Greene2023UNCOVERz5, Maiolino2023JADES.Mighty, Harikane2023JWST/NIRSpecProperties}. In addition, deeper data can reveal if there are broad components in other emission lines besides Balmer lines, or at least set limits on the strength of the AGN and the extent of dust extinction. High resolution spectroscopy will identify which emission lines are AGN or star-formation powered, and will also resolve doublets for better estimates of line luminosities, and in turn physical properties \citep[e.g.][]{Kocevski2023HiddenCEERS}. In particular, accurate estimates of the metallicities of LRDs could be extremely informative. IFS data, as in \citet{Parlanti2023GA-NIFS:z=4.76}, would help map the relative spatial extent of broad \ha and narrow \oiii. Further, since time variable measurements provide much more accurate SMBH properties than single-epoch derivations, observing LRDs at intervals to detect potential  variability over time would provide better constraints on AGN powering of these systems.

As a new class of objects, LRDs would benefit from their own template for SED fitting so that we may group similar objects under a new, more appropriate classification. The NIRSpec spectrum presented here for \lrd has very good SNR and may be used as a template\footnote{https://github.com/gbrammer/eazy-photoz/tree/master/templates/sfhz} for future LRD samples. 

\section{Conclusions}
We have presented the \emph{JWST} NIRCam images and NIRSpec spectrum of \lrd, a gravitationally lensed compact source with an unusual ``V-shaped'' continuum. We fit both the morphology and spectrum using various models, describing outflows, mergers, dust-obscured star-formation, AGN activity, and AGN dust torus emission. The best-fit to the spectrum is obtained by assuming two distinct components for the UV and optical/NIR sides of the spectrum, with different power law slopes and dust extinctions. We also find that while nearly all emission lines fit well with a Gaussian of width under 1000\,km\,s$^{-1}$, the \ha line shows evidence of AGN-broadening with a velocity width of $4300\pm300$\,km\,s$^{-1}$, indicating that \lrd appears to be an LRD with one of the highest inferred SMBH masses so far. Our modelling favours a scenario where the UV continuum arises from a star-forming region with low obscuration, the narrow emission lines arise either from star-formation and/or possibly an AGN NLR, and the optical/NIR continuum and broad line emission arise from the AGN and its surrounding BLR. The morphology, extended in the UV, and more compact towards the NIR, supports the domination of extended star-formation in the UV and compact AGN emission towards the NIR. We therefore conclude that the system is a highly obscured AGN within a less obscured star-forming host galaxy. The exhibition of features of both AGN (broad \ha line) and star-formation (spatially extended emission) in \lrd may indicate the current stage of evolution of a young AGN that will eventually grow and dominate over the host galaxy flux \citep{Fujimoto2022ADawn,Matthee2023EIGER.JWST}. LRDs in general may evolve into the AGN-dominated, bright blue quasars we see at lower redshifts, requiring a black hole mass increase of nearly two orders of magnitude. This would entail near-Eddington or super-Eddington accretion rates or mergers over the next few billion years. We proposed that LRDs may be low-metallicity type~2 AGN with very low dust-to-gas ratios, viewed \emph{through} their dust torus, giving rise to the direct detectability of their AGN in restframe red and near-infrared light.

\section{Acknowledgments}
The data products presented herein were retrieved from the DAWN \emph{JWST} Archive (DJA), an initiative of the Cosmic Dawn Center. The Cosmic Dawn Center is funded by the Danish National Research Foundation under grant DNRF140.

\vspace{5mm}

\bibliographystyle{aa} 
\bibliography{references}

\end{document}